# Giant Chern number of a Weyl nodal surface without upper limit

J.-Z. Ma[1,2,3,4*†], S.-N. Zhang[5,6†], J. P. Song[7,8], Q.-S. Wu[5,6], S. A. Ekahana[2], M. Naamneh[2,9], M. Radovic[2], V. N. Strocov[2], S.-Y. Gao[10], T. Qian[10,11], H. Ding[10,11,12], K. He[13], K. Manna[14,15], C. Felser[14], N. C. Plumb[2], O. V. Yazyev[5,6], Y.-M. Xiong[16*], M. Shi[2*]

**Affiliations:**

[1]*Department of Physics, City University of Hong Kong, Kowloon, Hong Kong, China*

[2]*Swiss Light Source, Paul Scherrer Institute, CH-5232 Villigen PSI, Switzerland*

[3]*City University of Hong Kong Shenzhen Research Institute, Shenzhen, China*

[4]*Hong Kong Institute for Advanced Study, City University of Hong Kong, Kowloon, Hong Kong, China*

[5]*Institute of Physics, École Polytechnique Fédérale de Lausanne, CH-10 15 Lausanne, Switzerland*

[6]*National Center for Computational Design and Discovery of Novel Materials MARVEL, École Polytechnique Fédérale de Lausanne (EPFL), CH-1015 Lausanne, Switzerland*

[7]*Anhui Key Laboratory of Condensed Matter Physics at Extreme Conditions, High Magnetic Field Laboratory, HFIPS, Anhui, Chinese Academy of Sciences, Hefei 230031, P. R. China*

[8]*University of Science and Technology of China, Hefei, Anhui 230026, China*

[9]*Department of Physics, Ben-Gurion University of the Negev, Beer-Sheva, 84105, Israel*

[10]*Beijing National Laboratory for Condensed Matter Physics and Institute of Physics, Chinese Academy of Sciences, Beijing 100190, China*

[11]*Songshan Lake Materials Laboratory, Dongguan, Guangdong, China*

[12]*CAS Center for Excellence in Topological Quantum Computation, University of Chinese Academy of Sciences, Beijing, 100049, China*

[13]*Department of Physics, Tsinghua University, Beijing 100084, P. R. China.*

[14]*Max Planck Institute for Chemical Physics of Solids, Dresden, D-01187, Germany*

[15]*Department of Physics, Indian Institute of Technology Delhi, Hauz Khas, New Delhi 110016, India*

[16]*Department of Physics, School of Physics and Optoelectronics Engineering, Anhui University, Hefei 230601, China*

*Corresponding to: junzhama@cityu.edu.hk, yxiong@hmfl.ac.cn, ming.shi@psi.ch

†These authors contributed equally to this work

Weyl nodes can be classified into zero-dimensional (0D) Weyl points (WPs), 1D Weyl nodal lines (WNL) and 2D Weyl nodal surfaces (WNS), which possess finite Chern numbers. Up to date, the largest Chern number of WPs identified in Weyl semimetals is 4, which is thought to be a maximal value for linearly crossing points in solids. On the other hand, whether the Chern numbers of nonzero-dimensional linear crossing Weyl nodal objects have one upper limit is still an open question. In this work, combining angle-resolved photoemission spectroscopy with density functional theory calculations, we show that the chiral crystal AlPt hosts a cube-shaped charged Weyl nodal surface which is formed by the linear crossings of two singly-degenerate bands. Different to conventional Weyl nodes, the cube-shaped nodal surface in AlPt is enforced by nonsymmorphic chiral symmetries and time reversal symmetry rather than accidental band crossings, and it possesses a giant Chern number $|C| = 26$. Moreover, our results and analysis prove that there is no upper limit for the Chern numbers of such kind 2D Weyl nodal object.

## I. Introduction

Weyl semimetals (WSMs) host singular degenerate nodes in the momentum space that are the sources or drains of Berry curvatures field [1–3]. Integrating the flux of this Berry curvature field on a 2D surface enclosing the Weyl degenerate node gives us a topological invariant quantity called the Chern number, whose value is analogous to a magnetic monopole charge. From the dimensionality, these Weyl nodes can exist as 0D Weyl point [1–3], 1D Weyl nodal line [4–6] (or loop, chain, etc.), 2D Weyl nodal surface [7–10] (or wall, plane, sphere, cube, etc.), as shown in Fig. 1(a). While massive theoretical and experimental efforts have been spent on electronic Weyl points [11–14] and Weyl nodal lines [4–6], the experimental identification of 2D electronic Weyl nodal surface has rarely been reported except some very recent works [15–17].

Weyl points associated with Chern number $|C|= 1$ or 2 have been identified in many materials [3,11,18–21] with broken time-reversal or inversion symmetry, where they are formed by the linear crossings of two bands at generic $k$ points. Further investigations of topological semimetals even revealed that the generalized Weyl fermions with Chern numbers higher than 2 can occur at time-reversal invariant momenta in some systems [22–24]. For example, spin-3/2 Rarita-Schwinger-Weyl (RSW) fermions and double spin-1 Weyl fermions with Chern numbers up to 4 have been identified in the CoSi family materials [22,25–29]. As no Chern number of Weyl points larger than 4 was reported based on experiments or density functional theory (DFT) calculations, it is thought that the Chern number could be limited up to 4 for single linearly crossed nodal point in solids [30–34]. The question about upper limit of Chern number is important as most topological properties are directly associated with it. However, the limit on the maximal Chern number of Weyl nodal point systems may not be directly applied to the less studied Weyl nodal surfaces, as we will demonstrate in this letter that there is no upper limit of Chern number for 2D Weyl nodal surface.

In this work, through symmetry analysis and using angle-resolved photoemission spectroscopy (ARPES), we show that the Chern number of a 2D Weyl nodal object can be significantly larger than 4. Specifically, we demonstrate that the chiral crystal AlPt hosts an enclosed 2D Weyl nodal cube (cube-shaped nodal surface) protected by combination of crystalline symmetries and time reversal symmetry with a Chern number as high as $|C|=26$. This Weyl nodal cube has two-fold band degeneracy formed by the linear crossings of two nondegenerate bands at the boundaries of the BZ. The Weyl nodal cube forms an enclosed cubic shape which exactly coincides with the BZ boundary. Moreover, the Berry curvature field of the Weyl nodal cube is absorbed by 0D unpaired Weyl points inside the BZ, - i.e., the Chern number of the Weyl nodal cube is opposite to the net Chern number of the Weyl points. Since no constraint that limits the number of 0D band crossings and net Chern number of the accidental WPs, our observation indicates that there is no upper limit for the Chern numbers of 2D Weyl nodal objects.

## II. Experimental and Calculation Details

Single crystals of AlPt were grown by using the melting method[25]. Firstly, the stoichiometric polycrystalline sample was obtained by arc melting the elements Al and Pt. Then, the crushed polycrystalline AlPt powder was loaded into an $Al_2O_3$ crucible and sealed in a Niobium tube. The reactant was heated up to 1873 K in the argon atmosphere and then cooled to 673 K with 0.5 K/min. Finally, the reactant was annealed at 1273 K for one week and cooled to the room temperature with 1 K/min. The obtained AlPt single crystals were characterized by the X-ray diffraction measurements (XRD).

ARPES measurements were performed at the SIS and ADRESS beamlines of the Swiss Light Source (PSI) [35,36], at DREAMLINE of the Shanghai Synchrotron Radiation Facility (SSRF), and at the beamlines UE112 PGM-2b-1^3 and UE112 PGM-2a-1^2 of the Berlin Electron Storage Ring Society for Synchrotron Radiation synchrotron (BESSY). The energy and angular resolutions were set to 5~100 meV and 0.1°, respectively. The experiments were conducted in a temperature range between 1 K and 20 K in a vacuum better than $2\times10^{-10}$ Torr.

The first-principles calculations have been performed using the Vienna ab initio simulation package (VASP) [37,38] within the GGA approximation. The cutoff energy of 520 eV was chosen for the plane wave basis. The lattice constants (4.866 $\dot{A}$) and atomic positions have been adopted from the ICSD database without carrying out any additional relaxation. The location of Weyl points and the calculation of their chirality have been performed using the open-source code WannierTools [39] based on the Wannier tight-binding model constructed using the Wannier90 code [40].

## III. Results and Discussion
### A. Crystal and electronic structure of AlPt

The chiral crystal AlPt has a cubic lattice structure belonging to the CoSi family with space group $P2_13$ (No.198) [22](Fig.1(c)). The cubic BZ and high-symmetry points are shown in Fig. 1(d). There are 12 symmetry operations associated with the three basis axes: two two-fold screw symmetry axes along the *z* and *x* directions, and one three-fold rotational symmetry axis along the (111) direction, respectively. The calculated band structure along high-symmetry lines is plotted in Fig. 1(e). Protected by the nonsymmorphic crystal symmetries and time-reversal symmetry, bands *N-3* and *N-2* are degenerate at all *k* points on the BZ boundary (black curves) and form a nodal cube which spans the energy range from around -1.2 to 0.2 eV across the Fermi level. Away from the BZ boundary, except at some discrete *k* points, the two bands are spin-polarized and nondegenerate bands (blue and green curves). While the unavoidable band crossing at the BZ centre (the Γ point) is protected by time-reversal symmetry, some accidental crossing points can also occur at generic *k* points (See Supplemental Material at table S1). These crossing points are the unconventional singular Weyl points, which do not pair with other Weyl nodes with opposite charge, but rather with the Weyl nodal cube on the BZ boundary [8]. As a consequence, no surface Fermi arc is needed to connect the surface projections of the Weyl points.

The non-trivial topological nature of the nodal cube can be quantified by calculating its Chern number. As these cubes are periodically connected with coincidence of the boundaries of BZs, it is impossible to find a surface that encloses the nodal cube for calculating the associated Chern number (Fig. 1(b)). We overcome this difficulty by choosing a closed gapped sphere-like path in the first BZ which is close enough to the boundaries of the BZ cube and encloses all the band crossings in the first BZ indicated by the dotted red lines in Fig. 1(b). From here, we find that the topological charge associated with this chosen surface is +13 in the enclosed volume, i.e. the net Chern number is +13 in the first BZ when the zone boundaries are excluded. As the net Chern number should be zero over the whole BZ, a topological charge of -13 at the BZ boundaries is required to absorb all the Berry curvature emerging from the band crossings inside the BZ. Due to the periodicity of reciprocal space, the boundaries between the first and the second BZs should also absorb the Berry curvature flowing into them from both sides. As a result, a total topological charge of -26 (Chern number) is required to reside on the nodal cube to neutralise the net Chern number of WPs in the 1[st] and 2[nd] BZs. When considering the Berry curvature field flowing from the third and fourth BZs to the corners or edges of the first BZ boundary, the absolute value of the Chern number of the cube could be larger than 26. However, such a value is incalculable. Under a premise that the density of charge $\rho$ on the nodal cube is finite, the problem is solved as shown below. The integration of the Berry curvature from the 3rd and 4th BZs to the 1st BZ cube is ($\int \rho \cdot \Delta S$), where $\Delta S$ is the infinitesimal area of the edges and corners, and $\rho$ is the density of charge with finite value. The integration area of infinitesimal edges or corners are zero, i.e., ($\int \rho \cdot \Delta S = 0$). Thus, we don't need to consider the Berry curvature from 3$^{rd}$ and 4$^{th}$ BZs to the first one under such premise.

### B. Observation of the Weyl nodal cube

To verify the band structure analysis from the DFT calculation, we performed a systematic ARPES measurements, focused on the boundaries of the bulk BZ. Figure 2(b) shows the Fermi surface (FS) map in the R-M-X plane, which is a face of the Weyl nodal cube. The photon energies for different $k_z$ are determined from photon energy dependent measurements (See Supplemental Material at Fig. S1 for more details). We observed circle-like FS pockets around the R points and small FS patches around the M points. The ARPES intensity in the R-M-X plane, as a function of energy relative to the Fermi level ($E_F$), is plotted in Fig. 2(c). The Weyl nodal cube is marked with white dotted lines which cross the $E_F$ to form the pieces of FS around the M point, indicating its non-trivial topology at the Fermi level. In the R-M-X plane, there are four types of high symmetry lines, i.e., R-M, X-M along the $x$ axis (X1-M), X-M along the $y$ axis (X2-M) and R-X. Although the X points along $k_x$, $k_y$, $k_z$ axes are equivalent in a BZ, the X1-M and X2-M lines are in fact not equivalent as the system has a nonsymmorphic screw symmetry, instead of the 4-fold rotational symmetry along the $k_x$, $k_y$ or $k_z$ axes. Figures 2(e-h) show the ARPES spectra along the high-symmetry lines. The observed band dispersions indicated by green arrows agree well with the Weyl nodal cube identified in the band structure calculations, which are shown with black lines in Figs. 2(i-l). The nontrivial nodal cube spans an energy range from around -1.2 eV to 0.2 eV. As the nodal cube crosses $E_F$ near the M points, it leads to the formation of nontrivial Fermi surfaces with nonzero Chern number. In previous studies on topological materials, it has been shown that the nonzero Berry curvature field at the Fermi level contributes to transport and magneto-transport properties as observed in the Hall effect and magnetic resistance experiments [1,41–44]. In the momentum space, the Chern number and Berry curvature behave like magnetic charge and magnetic field, which can affect the transport behaviour of electrons. The 2D Weyl nodal cube has similar nontrivial topological characters as the 0D Weyl points, but geometrically it is more extended in 3D BZs. The unique geometry of 2D Weyl nodal cube would be reflected in the anisotropy of transport property, as well as Landan levels and thus the chiral anomaly. It is expected that the nodal cube would contribute to low-energy excitations and affect the transport properties, but could have certain differences to the WSM with 0D nodes.

### C. The Nielsen-Ninomiya no-go theorem

The enclosed Weyl nodal cube in AlPt makes this material very special when considering the well-known Nielsen-Ninomiya no-go theorem [45,46]. The theorem stipulates an equal number of oppositely charged Weyl points, such that the total Chern number is zero, meaning that Weyl points must appear in pairs. The no-go theorem is only proven, however, in the case of 0D Weyl nodes (Weyl points), while, its applicability to 2-D Weyl nodal surfaces has not been established. This no-go theorem can be visualized as follows [8]: We choose a 2D-slice in the BZ, which is perpendicular to the surface BZ as depicted by the light green and light red planes in Fig. 1a (i), and calculate the Chern number on this slice. Moving the slice in the reciprocal space will cause the Chern number changes (e.g. from zero to an integer) once it crosses a Weyl node (the blue Weyl node in Fig. 1a (i)). Due to the periodicity of the reciprocal space, the Chern number on the slice has to return to its initial value as it passes through the starting point in the BZ. Therefore, it is reasonable to require the existence of another oppositely charged Weyl node to neutralise the initial change/jump (the red Weyl node in Fig. 1a (i)). The nonzero Chern number on the slice (the light red plane in Fig. 1a (i)) reflects the related number of chiral surface states at the surface BZ projection, thus forming the Fermi arcs connecting the projections of the bulk Weyl points. However, this scenario is circumvented if there is a Weyl nodal cube on the BZ boundary, as the Chern number on the 2D slice is ill-defined without fully gapped bulk structure. In this case, the Weyl nodal points inside the BZ do pair with the Weyl nodal cube, and the surface Fermi arcs are no longer required to connect the projections of the Weyl points (Fig. 1a (iii)). Instead, all the extra Berry curvature starting or ending at a Weyl point can be absorbed by the Weyl nodal cube (Fig. 3(a)).

### D. Unpaired Weyl points inside the BZ

In the following, we support the claim of unpaired Weyl points by acquiring ARPES spectra in the Γ-X-M plane (Fig. 3(b)). The FS map and ARPES intensity as a function of energy in the Γ-X-M plane are displayed in Figs. 3(c,d). The unpaired Weyl point at the BZ centre (yellow dot in Fig.3(d)) is surrounded by the Weyl nodal cube (red dotted lines in Fig. 3(d)) and is found to carry a Chern number of +1 according to our calculation. Unlike conventional WPs that appear in pairs and require no extra protection by symmetries, the unpaired WP at the Γ point is protected by time-reversal symmetry. Figures 3(e,f,i) show the ARPES spectra along Γ-X acquired with both soft X-ray and ultraviolet X-ray [47]. The observed band dispersions agree very well with the calculated band structure along Γ-X (Fig. 3(g,h,j)). The RSW node and the singular Weyl point (SWP) is clearly observed at Γ point (Fig. 3(e,f)), and the linear crossing of bands N-3 and N-2 at the X point forming the WNC is clearly visible in Fig. 3(f,i). Figure 3(k) shows the ARPES spectra along F-M direction indicating that the bands *N-3* and *N-2* do cross Fermi level near the M point, and it is reproduced by the calculation in Fig. 3(l).

As our calculation suggests a net charge of +13 inside the BZ, we conclude that there exists other accidental WPs on the generic $k$ points to compensate the total net charge. By searching for the band crossings thoroughly in the entire BZ via first principle calculations, we find that there are additional 3 types of WPs on high symmetry planes, i.e. 12 W1 points with Chern number +1; 12 W2 with Chern number +1, and 12 W3 with Chern number -1. In total, we propose 37 WPs, including the WP at the Γ point, with a net Chern number +13 enclosed by the Weyl nodal cube. The coordinates, chirality, and energies of the Weyl points are listed in Supplemental Material at table S1. In a summary, there are 37 0D Weyl points and one 2D Weyl nodal cube formed by band N-3 and N-2, and no band crossing between these two bands at all the other $k$ points. The 37 Weyl points positions are displayed in one 3D BZ in Supplemental Material at Fig. S2. To facilitate the understanding of the topological connection between the unpaired Weyl points and the Weyl node cube, we plotted the in-plane component of the Berry curvature in the $k_z$=0 plane (See Supplemental Material at Fig. S3). It indicates that the extra Berry curvature field starting from (ending at) the unpaired Weyl points flows into (comes out from) the Weyl nodal cube on the BZ boundary.

Only the Weyl point on the BZ centre and the 2D Weyl nodal cube are protected by symmetries, and the 36 Weyl nodes on generic k points are formed accidentally by band crossings. On the other hand, it is known that the detailed band dispersion can be tuned by varying lattice and other physical parameters, thus the total Chern number may be manipulated by creating or removing the band crossings inside the BZ via mechanical, chemical, optical, or other means of perturbation. Accordingly, the net Chern number of the WPs will follow the formula $(1+6m+12n+24q)$ where $m$, $n$, $q$ are integer corresponding to the Weyl points on high symmetry lines, planes, and generic $k$ points, respectively. Since no theory restricts the number of accidental band crossings with finite Chern number in a BZ, our results imply that the absolute value of the Chern number possessed by the Weyl nodal cube is not limited to 26, i.e. there is no upper limit on the Chern number for such kind of Weyl nodal surface. It should be noted that in the previous study of PtGa [17], we used the formula 1+24m-12n (m, n are either 1 or 0 to count the number of possible accidental Weyl points in generic *k* points and high symmetry planes) to describe the net charge of the Weyl points. That work focuses on topological protection of the net chiral charge and try to find the lowest Chern number without discussing the upper limit.

## IV. Conclusion

In conclusion, with the joint theoretical and experimental efforts we have uncovered the existence of a Weyl nodal cube-shaped surface with giant topological charge of -26 formed by two linearly singly-degenerated band crossings. This finding enlarges our understanding of the topological nature of Weyl nodes and should prompt further inquiry into whether there exists an upper limit for the Chern number of doubly-degenerate Weyl nodes. Our results and analysis indicate that there is no upper limit for the Chern number of 2D Weyl nodal surfaces, in contrast to the seemingly maximal value of 4 for 0D Weyl points. Meanwhile, this Weyl nodal cube with the giant Chern number that crosses the Fermi level will contribute to the low-energy excitations, which promises to become an important topic for future transport or magneto transport studies. It will likewise stimulate further explorations for novel topological materials, where new topological physics phenomena are waiting to be discovered.

## Acknowledgments

We acknowledge T. L. Yu, Y. B. Huang, E. Rienks, M, Krivenkov, A. Varykhalov for help during the ARPES experiments. **Funding:** This work was supported by the National Natural Science Foundation of China (12104379), the NCCR-MARVEL funded by the Swiss National Science Foundation, the Sino-Swiss Science and Technology Cooperation (Grant No. IZLCZ2-170075), and the Swiss National Science Foundation under Grant. No. 200021-188413. M.R. and J.Z.M. were supported by project 200021_182695 funded by the Swiss National Science Foundation. J.Z.M is supported by City University of Hong Kong through the start-up project (Project No. 9610489), and by City University of Hong Kong Shenzhen Institute. S.A.E acknowledges the support from NCCR-MARVEL funded by the Swiss National Science Foundation and the European Union's Horizon 2020 research and innovation programme under the Marie Skłodowska-Curie grant agreement No. 701647. M.N. was supported by Swiss National Science Foundation project 200021_159678. H.D. and T.Q. acknowledge financial support from the Ministry of Science and Technology of China (2016YFA0401000 and 2016YFA0300600), the National Natural Science Foundation of China (U1832202), the Chinese Academy of Sciences (QYZDB-SSW-SLH043, XDB33000000, and XDB28000000). K.M., and C.F. acknowledge financial support from European Research Council Advanced Grant No. (742068) "TOP-MAT," European Union's Horizon 2020 research and innovation programme (grant No. 824123 and 766566) and Deutsche Forschungsgemeinschaft (Project-ID 258499086 and FE 633/30-1). Y.M.X acknowledges financial support from the National Key Research and Development Program of China (Grant No. 2016YFA0300404), and the Collaborative Innovation Program of Hefei Science Center, CAS(Grant No. 2019HSC-CIP007).

Takahashi, Y. Ando, and T. Sato, *Observation of Chiral Fermions with a Large Topological Charge and Associated Fermi-Arc Surface States in CoSi*, Phys. Rev. Lett. **122**, 076402 (2019).

[26] Z. Rao, H. Li, T. Zhang, S. Tian, C. Li, B. Fu, C. Tang, L. Wang, Z. Li, W. Fan, J. Li, Y. Huang, Z. Liu, Y. Long, C. Fang, H. Weng, Y. Shi, H. Lei, Y. Sun, T. Qian, and H. Ding, *Observation of Unconventional Chiral Fermions with Long Fermi Arcs in CoSi*, Nature **567**, 496 (2019).

[27] D. S. Sanchez, I. Belopolski, T. A. Cochran, X. Xu, J. X. Yin, G. Chang, W. Xie, K. Manna, V. Süß, C. Y. Huang, N. Alidoust, D. Multer, S. S. Zhang, N. Shumiya, X. Wang, G. Q. Wang, T. R. Chang, C. Felser, S. Y. Xu, S. Jia, H. Lin, and M. Z. Hasan, *Topological Chiral Crystals with Helicoid-Arc Quantum States*, Nature **567**, 500 (2019).

[28] N. B. M. Schröter, D. Pei, M. G. Vergniory, Y. Sun, K. Manna, F. de Juan, J. A. Krieger, V. Süss, M. Schmidt, P. Dudin, B. Bradlyn, T. K. Kim, T. Schmitt, C. Cacho, C. Felser, V. N. Strocov, and Y. Chen, *Chiral Topological Semimetal with Multifold Band Crossings and Long Fermi Arcs*, Nat. Phys. **15**, 759 (2019).

[29] M. Yao, K. Manna, Q. Yang, A. Fedorov, V. Voroshnin, B. Valentin Schwarze, J. Hornung, S. Chattopadhyay, Z. Sun, S. N. Guin, J. Wosnitza, H. Borrmann, C. Shekhar, N. Kumar, J. Fink, Y. Sun, and C. Felser, *Observation of Giant Spin-Split Fermi-Arc with Maximal Chern Number in the Chiral Topological Semimetal PtGa*, Nat. Commun. **11**, 2033 (2020).

[30] C. Fang, M. J. Gilbert, X. Dai, and B. A. Bernevig, *Multi-Weyl Topological Semimetals Stabilized by Point Group Symmetry*, Phys. Rev. Lett. **108**, 266802 (2012).

[31] J. Cano, B. Bradlyn, and M. G. Vergniory, *Multifold Nodal Points in Magnetic*16Takahashi, Y. Ando, and T. Sato, *Observation of Chiral Fermions with a Large Topological Charge and Associated Fermi-Arc Surface States in CoSi*, Phys. Rev. Lett. **122**, 076402 (2019).

[26] Z. Rao, H. Li, T. Zhang, S. Tian, C. Li, B. Fu, C. Tang, L. Wang, Z. Li, W. Fan, J. Li, Y. Huang, Z. Liu, Y. Long, C. Fang, H. Weng, Y. Shi, H. Lei, Y. Sun, T. Qian, and H. Ding, *Observation of Unconventional Chiral Fermions with Long Fermi Arcs in CoSi*, Nature **567**, 496 (2019).

[27] D. S. Sanchez, I. Belopolski, T. A. Cochran, X. Xu, J. X. Yin, G. Chang, W. Xie, K. Manna, V. Süß, C. Y. Huang, N. Alidoust, D. Multer, S. S. Zhang, N. Shumiya, X. Wang, G. Q. Wang, T. R. Chang, C. Felser, S. Y. Xu, S. Jia, H. Lin, and M. Z. Hasan, *Topological Chiral Crystals with Helicoid-Arc Quantum States*, Nature **567**, 500 (2019).

[28] N. B. M. Schröter, D. Pei, M. G. Vergniory, Y. Sun, K. Manna, F. de Juan, J. A. Krieger, V. Süss, M. Schmidt, P. Dudin, B. Bradlyn, T. K. Kim, T. Schmitt, C. Cacho, C. Felser, V. N. Strocov, and Y. Chen, *Chiral Topological Semimetal with Multifold Band Crossings and Long Fermi Arcs*, Nat. Phys. **15**, 759 (2019).

[29] M. Yao, K. Manna, Q. Yang, A. Fedorov, V. Voroshnin, B. Valentin Schwarze, J. Hornung, S. Chattopadhyay, Z. Sun, S. N. Guin, J. Wosnitza, H. Borrmann, C. Shekhar, N. Kumar, J. Fink, Y. Sun, and C. Felser, *Observation of Giant Spin-Split Fermi-Arc with Maximal Chern Number in the Chiral Topological Semimetal PtGa*, Nat. Commun. **11**, 2033 (2020).

[30] C. Fang, M. J. Gilbert, X. Dai, and B. A. Bernevig, *Multi-Weyl Topological Semimetals Stabilized by Point Group Symmetry*, Phys. Rev. Lett. **108**, 266802 (2012).

[31] J. Cano, B. Bradlyn, and M. G. Vergniory, *Multifold Nodal Points in Magnetic*
16

*Density Waves*, Phys. Rev. Lett. **109**, 086401 (2012).

[48] P. Zhang, P. Richard, T. Qian, Y. M. Xu, X. Dai, and H. Ding, *A Precise Method for Visualizing Dispersive Features in Image Plots*, Rev. Sci. Instrum. **82**, 043712 (2011).

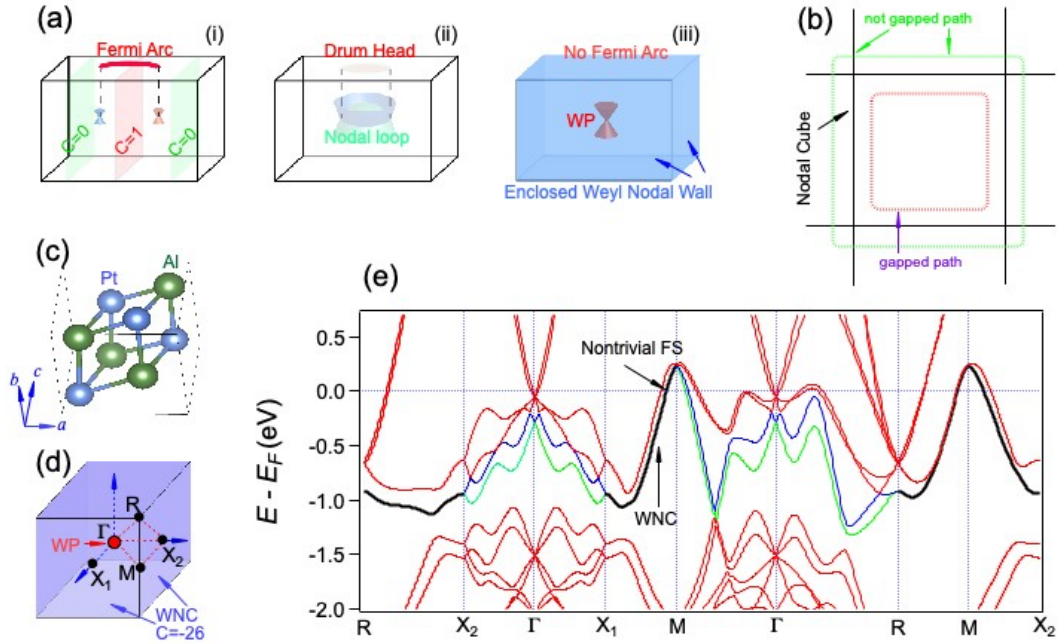

**Figure 1. Classification of Weyl semimetals (WSMs) with different dimensional nodal objects and the Weyl nodal cube in AlPt.** (a), Schematics of WSMs with point nodes, line nodes, sphere nodes, and wall nodes, respectively. (b), Path enclosing and enclosed by the nodal cube. (c), Conventional unit cell of AlPt crystal. (d), Bulk BZ of AlPt with some high-symmetry points labeled. The unpaired Weyl point (WP) in the BZ center is enclosed by the topologically charged Weyl nodal cube (WNC) on the boundary of the BZ. No surface Fermi arc connects the surface projection of this singular WP. (e), Band structure along high-symmetry lines in AlPt. The black bands indicate the charged Weyl nodal cube (WNC).

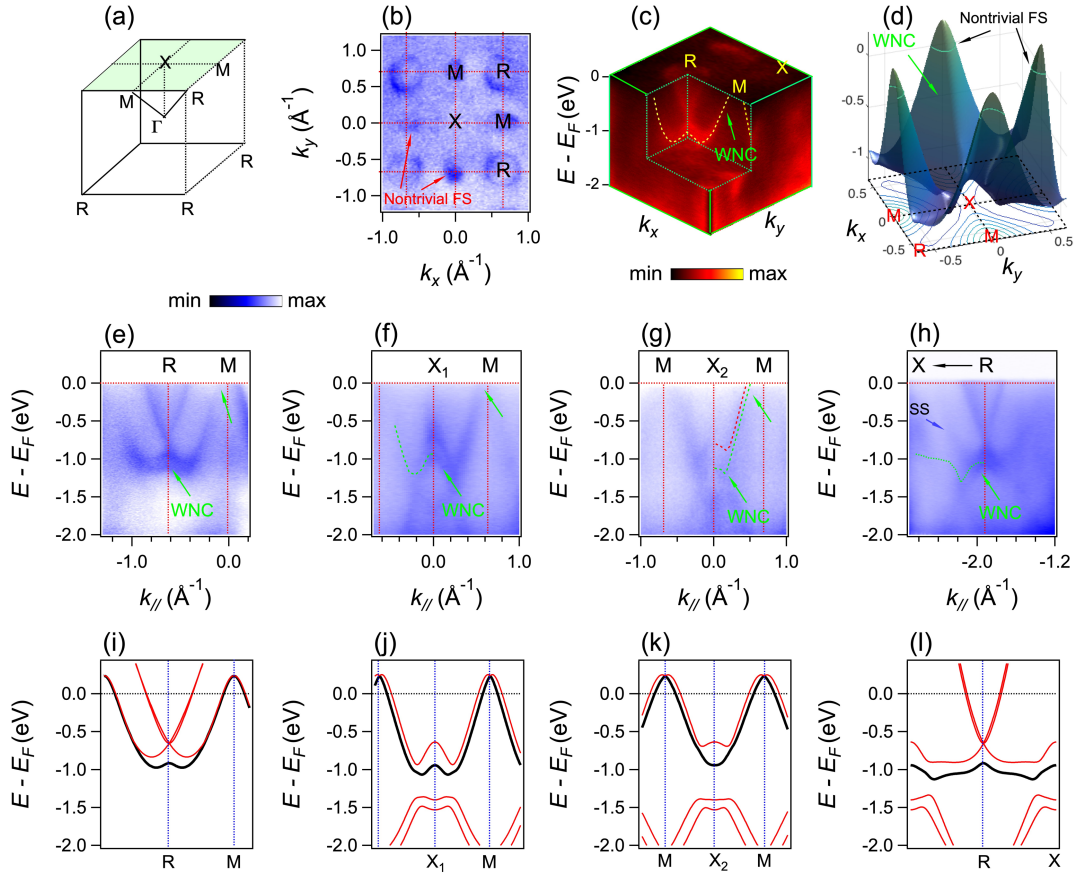

**Figure 2. Weyl nodal cube on one surface of the cubic BZ boundary.** (a), Bulk BZ of AlPt with high-symmetry points labeled. (b), FS map in the MXR plane marked as light green shading in a, acquired with $h\nu = 122$ eV. (c), ARPES intensity plot in the R-X-M plane as a function of energy relative to the Fermi level. The WNC is marked with white dotted lines, which are double degenerate states protected by symmetries at all $k$ points on the R-X-M plane. (d), Calculated 3D Weyl nodal cube along the X-M-R plane. (e-h), ARPES spectra along high symmetry lines on the BZ boundary. The green arrows indicate the WNC which crosses the Fermi level near the M point. (i-l), Calculated band dispersions along the high-symmetry lines on the BZ boundaries. The black curves denote the WNC.

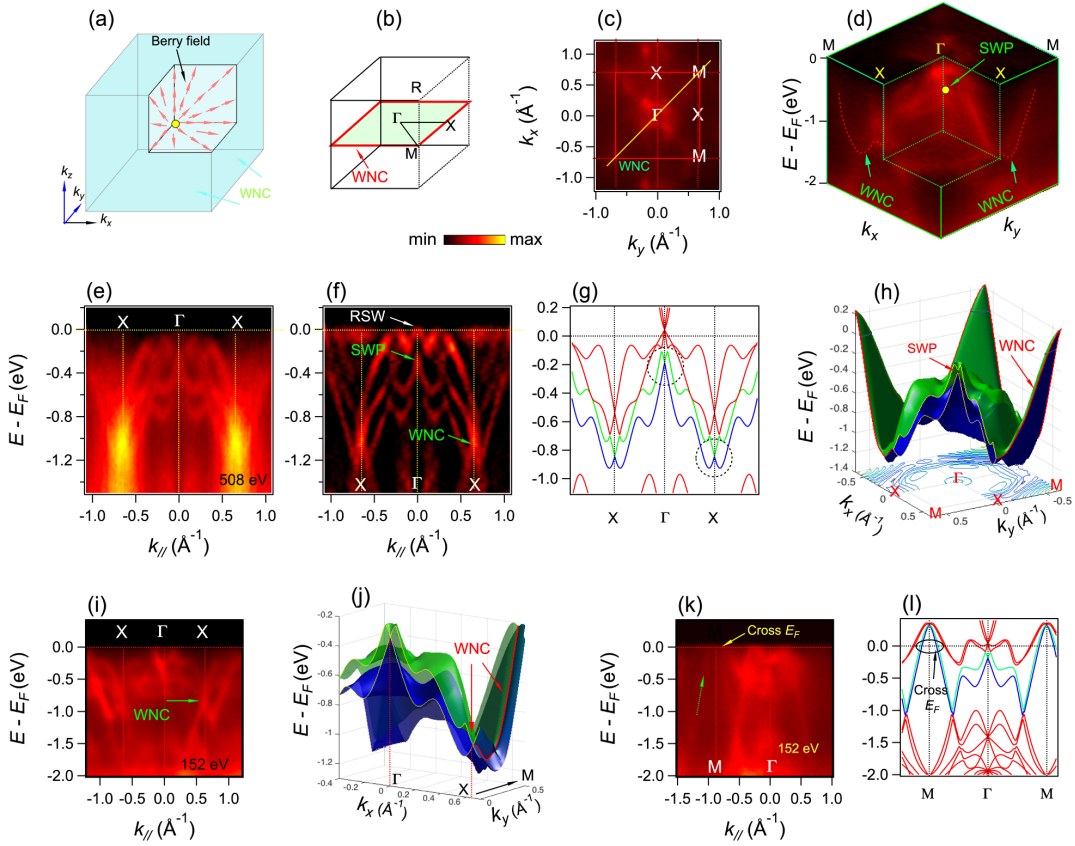

**Figure. 3. Unpaired WP inside the WNC.** (a), Schematic illustration of an unpaired Weyl point in the BZ center enclosed by the topologically charged WNC on the BZ boundary. (b), Bulk BZ of AlPt with some high symmetry points labeled. The light green plane indicates the Γ-X-M plane, and the red boundary indicates the WNW as one cross-section between the Γ-X-M plane and the WNC. (c), ARPES FS map in the Γ-X-M plane. (d), The 3D ARPES intensity plot in the Γ-X-M plane as a function of energy relative to the Fermi level. The unpaired WP at the Γ point and the WNWs on the BZ boundary are marked with the filled yellow circle and red dotted lines, respectively. (e,f), ARPES spectra acquired with $h\nu$ = 508 eV along the Γ-X line and its curvature intensity plot [48]. (g), Calculated band structure along the Γ-X line. (h), Calculated 3D band structure of bands *N-3* and *N-2* in the Γ-X-M plane. The single Weyl point (SWP) at the BZ center and the Weyl nodal cube (WNC) on the boundary are indicated by red arrows. (i), The ARPES spectra along the Γ-X line recorded with $h\nu$ = 152 eV, which shows the Weyl nodal cube band crossing at X point. (j), Zoomed-in view of the 3D band structure

in the vicinity of Γ-X line. (k), ARPES spectra along the Γ-M line indicating that the bands *N-3* and *N-2* do cross the Fermi level near the M point. (l), Calculated band structure along the Γ-M line.